# Accurate Estimation of Quantitative Trait Locus Effects with Epistatic by Improved Variational Linear Regression

**Dong Zijian [1], Wang Jingzhuo [1], Wang Zhongming [2]***

1. School of Electronic Engineering, Huaihai Institute of Technology, Lianyungang, Jiangsu, China;
2. Department of Mathematics and Statistics, Florida International University, FL, USA
* Author to whom correspondence should be addressed, E-Mail: dzjian@126.com, +8613003463463

**Abstract**

Bayesian approaches to variable selection have been widely used for quantitative trait locus (QTL) mapping. The Markov chain Monte Carlo (MCMC) algorithms for that aim are often difficult to be implemented for high-dimensional variable selection problems, such as the ones arising in epistatic analysis. Variational approximation is an alternative to MCMC, and variational linear regression (VLR) is an effective solution for the variable selection problems, but lacks accuracy in some QTL mapping problems where there are many more variables than samples. In this paper, we propose an effective method with aim to improve the accuracy of VLR in the case of above by dynamically reducing components (variable or markers) with known effects (zero or fixed). We show that the proposed method can greatly improve the accuracy of VLR with little increase in computational cost. The method is compared with several other variational methods used for QTL mapping, and simulation results show that its performance is higher than those methods when applied in high-dimensional cases.

**1. Introduction**

The goal of quantitative trait loci (QTLs) mapping and association studies is to identify certain regions of the genome that contain genes involved in specifying a quantitative trait, and to estimate the genetic effects of these loci. The relationship between the genetic effects of QTLs and the phenotypic value of quantitative traits can be described by a linear model [1, 2]. There are usually a large number of markers across the whole genome, and most of the markers have very little or even no effect on the phenotypes. So the model is sparse, and in most cases, the number of markers or variables is bigger than the sample size, especially when interactions among markers are considered. Therefore, the model is oversaturated and usually solved by shrinkage or variable selection methods.

The variable selection procedure for QTLs mapping can be seen as one of deciding which subset of variables have effects on phenotypes, and identifying out all possible effects of those markers. There are many ways of variable selection. For instance, the indicator selection uses a spike (probability distribution of variables not included in model) and slab (probability distribution of variables included in model) prior to indicate whether each variable is included in one model. Such methods usually adopt Gibbs sampling or importance sampling to calculate the weighted average as the inferred results [3-5].

Shrinkage methods do not use indicators to induce the sparseness of variables, but instead specify *a priori* directly on variables to approximate the "slab" and "spike" shapes and use a penalty function to shrink most variables toward zeros. Early Bayesian shrinkage methods employed Markov chain Monte Carlo (MCMC) techniques to infer the models [6], which were computationally intensive. Xu et al. proposed an empirical Bayes method that used a carefully chosen prior distribution for the variables, which required much less computation [7]. Furthermore, Cai et al. developed a fast empirical Bayesian LASSO algorithm [8], which capitalized the idea of relevance vector machine (RVM) to shrink some unimportant marker effects to zeros by using hierarchical sparsity-inducing priors [9]. Shrinkage methods can be seen as adaptive in the sense that the degree of sparseness of the model can be adjusted by changing the hyper parameters of a prior distribution [10].

Variational approximation techniques have also been extensively studied to solve the linear model [4, 11-14]. These methods can provide an analytical approximation to the posterior probability for the model. But when applied in a case such as QTL mapping with large-scale variables and a relatively small number of samples, these methods are either imprecise, or computationally intensive. Among these, the variational linear regression (VLR) is a terse method [15], but its accuracy is not good enough for large-scale QTL mapping problems. In this paper, we use the basic variational inference technology to infer the linear models, and then propose a novel method to improve the performance by reducing components (variables) in VLR dynamically (RCVLR). The proposed method is simple, effective, and accurate, especially in the case where there are many more variables than observations. Simulations show that RCVLR has satisfied speed and accuracy compared with other variational approximation methods. The idea of deleting some elements with very small effects from the model was also appeared in [16], but the main aim in it was to avoid numerical break-downs. Because the values of elements were related to the diagonal of the variance-covariance matrix, thus it would be singular when the elements value approached zeros.

This paper is organized as follows: we firstly introduce the variational inference of the linear model, and then propose the variable reduction technique for the linear model. In example analysis part, we demonstrate the effectiveness of the method through four examples by comparing with other algorithms, using the North American Barley data [17], the simulated data from Xu [2], and some simulated data sets with different variable sizes and different sample sizes. Finally, we briefly discuss the features of our methods and several other existing variational methods.

## 2. Methods
### 2.1 Linear regression model for multiple QTL mapping

Following standard practice, an interested phenotype is modeled as a linear combination of the genotypes plus residual noise[18]

$$y_j = \beta_0 + \sum_{k=1}^{p} x_{jk}\beta_k + e_j. \tag{1}$$

where $y_j$ is the phenotypic value of the $j^{\text{th}}$ individual, $x_{jk}$ is the genotypic value of individual $j$ at marker $k$, $\beta_0$ is the population mean, $\beta_k$ is the effect of the $k^{\text{th}}$ marker, and $e_j$ is the measurement error of individual $j$. Considering the epistatic effects, and defining the interaction effect of the maker pair $(u,v)$ as $\beta_{uv}$, we extend model (1) as

$$y_j = \beta_0 + \sum_{k=1}^{p} x_{jk}\beta_k + \sum_{u<v}^{p} x_{ju}x_{jv}\beta_{uv} + e_j \qquad (2)$$

Now, considering all elements, and assuming that the predictors and outcomes have been centered, we omit the intercept terms in the model and write the model (2) as

$$\mathbf{Y} = \mathbf{X}\boldsymbol{\beta} + \mathbf{e} \qquad (3)$$

where $\mathbf{Y} = (y_1, y_2, ..., y_N)^T$, $\mathbf{X}$ is a $N \times N_p$ matrix, and $\boldsymbol{\beta}$ is a $N_p \times 1$ vector with $N_p = p + \binom{p}{2} = \frac{p(p+1)}{2}$.

## 2.2 Variational approximation for linear models

Variational techniques have been extensively studied to infer the linear regression models [4, 11, 15]. In this section, we re-derive the basic variational infer technique for the linear regression model, and then give our variable reduction technique in Section 2.3.

In QTL mapping problems, $\boldsymbol{\beta}$ is sparse, which means that most entries in $\boldsymbol{\beta}$ are zeros, therefore we assume that $\boldsymbol{\beta}$ satisfies Gaussian distribution, $\boldsymbol{\beta} \sim N\left(0, \frac{1}{\alpha}\mathbf{I}\right)$, and $\alpha \sim Gamma(a_0, b_0)$, where $a_0, b_0$ are two hyper parameters. All residual errors are assumed to be independent and identically distributions [15], so $\mathbf{e} \sim N(0, \sigma^2 \mathbf{I})$.

The likelihood function is:

$$p(\mathbf{Y}|\boldsymbol{\beta}, \mathbf{X}) \sim \prod_{i=1}^{N} p(y_i | \boldsymbol{\beta}, \mathbf{X}_i), \qquad (4)$$

where $\mathbf{X}_i$ is the $i^{th}$ row of $\mathbf{X}$, and

$$p(y_i | \boldsymbol{\beta}, \mathbf{X}_i) \sim N(y_i | \mathbf{X}_i \boldsymbol{\beta}, \sigma^2). \qquad (5)$$

The joint distribution is

$$p(\mathbf{Y}, \boldsymbol{\beta}, \alpha) = p(\mathbf{Y}|\boldsymbol{\beta}, \alpha) p(\boldsymbol{\beta}|\alpha) p(\alpha). \qquad (6)$$

The variational approximation inference is to find a tractable distribution that is close to the target posterior. According to the approximate factorization theory [19], we have

$$p(\boldsymbol{\beta}, \alpha | \mathbf{X}, \mathbf{Y}) \approx q(\boldsymbol{\beta}, \alpha) = q(\boldsymbol{\beta}|\alpha)q(\alpha). \qquad (7)$$

We first calculate the approximation of the posterior probability of $\alpha$. According to the variational approximation theory, $\ln q(\alpha)$ is the expectation of the logarithm of the joint distribution, taking over all variables in $\boldsymbol{\beta}$.

$$\begin{aligned}\ln q(\alpha) &= E_{\boldsymbol{\beta}}\left[\ln p(\mathbf{Y},\boldsymbol{\beta},\alpha)\right]+const. \\ &= \left(a_0+N_p/2-1\right)\ln\alpha-\alpha\left(b_0+\frac{1}{2}E\|\boldsymbol{\beta}\|^2\right)+const.\end{aligned} \quad (8)$$

So, $q(\alpha) \sim Gamma(\alpha\,|\,a_n,b_n)$, and

$$\begin{cases} a_n = a_0 + N_p/2 \\ b_n = b_0 + \frac{1}{2}E\|\boldsymbol{\beta}\|^2 \end{cases} \quad (9)$$

where $E\|\boldsymbol{\beta}\|^2 = \|\boldsymbol{\mu}_n\|^2 + tr(\mathbf{S}_n)$, $\boldsymbol{\mu}_n, \mathbf{S}_n$ are the mean vector and variance matrix of the posterior distribution of $\boldsymbol{\beta}$ respectively, and will be calculated later.

We use the variational theory again to estimate the posterior probability of $\boldsymbol{\beta}$.

$$\begin{aligned}\ln q(\boldsymbol{\beta}) &= E_{\alpha}\left[p(\mathbf{Y},\boldsymbol{\beta},\alpha)\right]+const. \\ &= \frac{-1}{2}\boldsymbol{\beta}^T\left(\frac{1}{\sigma^2}\mathbf{X}^T\mathbf{X}+E(\alpha)\mathbf{I}\right)\boldsymbol{\beta}-\frac{1}{\sigma^2}\sum_i y_i\mathbf{X}_i\boldsymbol{\beta}+const.\end{aligned} \quad (10)$$

We can see that $q(\boldsymbol{\beta}) \sim N(\boldsymbol{\beta}\,|\,\boldsymbol{\mu}_n,\mathbf{S}_n)$, and

$$\begin{cases} \mathbf{S}_n = \left(\frac{1}{\sigma^2}\mathbf{X}^T\mathbf{X}+\frac{a_n}{b_n}\mathbf{I}\right)^{-1} \\ \boldsymbol{\mu}_n = \frac{1}{\sigma^2}\mathbf{S}_n\mathbf{X}^T\mathbf{Y} \end{cases} \quad (11)$$

There is a matrix inversion operation in (11). Since $\frac{1}{\sigma^2}\mathbf{X}^T\mathbf{X}+\frac{a_n}{b_n}\mathbf{I}$ is a $N_p \times N_p$ matrix and usually $N_p$ is much greater than $N$, The Woodbury matrix identity can be used to reduce the computational complexity of $\mathbf{S}_n$.

$$\mathbf{S}_n = \frac{b_n}{a_n}\mathbf{I} - \left(\frac{b_n}{a_n}\right)^2 \mathbf{X}^T\left(\sigma^2\mathbf{I}+\frac{b_n}{a_n}\mathbf{X}\mathbf{X}^T\right)^{-1}\mathbf{X} \quad (12)$$

Here, we only need to calculate the inversion of a matrix whose size is $N \times N$, which is much smaller than $N_p \times N_p$.

(9) and (11) are combined to constitute a complete iterative process. This iterative process should be executed with a sufficient number of times to reach a steady state. A simple stopping condition for the steady state is to test the numerical differences between two inferred parameters in two successive iterations, i.e. $S = \|\boldsymbol{\mu}_n - \boldsymbol{\mu}_{n-1}\|^2 \pm$. If $S$ is small enough (say smaller than $\xi$), the iteration is considered to reached a steady state. The choice of $\xi$ influences the algorithm accuracy. The smaller $\xi$ is, the higher the accuracy is, at the cost of more iterations and more computational time.

## 2.3 Reduce the size of linear model to improve accuracy

When there are many more variables than samples, the variational approximation above has poor performance. The reason is that VLR can not regress any variable to zero, and all the inferred values are close to zeros. However, from a positive point of view, we believe that the inferred value for each variable can reflect its probability (likelihood) of zero; therefore we can use the inferred results of VLR to delete some variables with very small absolute values. After deletion, we use VLR again to infer the parameters of the reduced-size model. In the case of a fixed number of measurements, VLR with smaller scale data undoubtedly has better inference performances. It may require several iterations to obtain better inferred results.

We first look at what elements should be removed from the model, that is, how to determine the threshold of deletion. When the size of variables is much larger than the size of samples, most of the parameter values derived by VLR are close to 0, so we need to set a relatively low threshold for deleting elements, say, $\eta_1 = 0.0001$. In general, we should choose the value of $\eta_1$ by the scale of problem. According to the results of the experiment, we set $\frac{1}{10 \times N_p} < \eta_1 < \frac{1}{5 \times N_p}$. After the first run of VLR, A lot of elements in the model would be deleted. With the reduced-sized model, we perform VLR again. As a whole, the parameter values derived by VLR will increase with the decrease of model size, and thus we need to increase the deletion threshold adaptively. In the algorithm, we set $\eta_h = \eta_1 + (h-1) \times \frac{6}{N_p}$ as the threshold for the $h$-th run of VLR.

After determining the deletion threshold, we examine how to remove components or elements. Let $\boldsymbol{\beta}$ is the original parameter vector, and $\mathbf{c}$ is corresponding binary-valued indicator vector of $\boldsymbol{\beta}$. If $c_k = 0$, then $\beta_k$ is small and can be deleted. If $c_k = 1$, then $\beta_k$ is relatively large and should be retained. Let $R = \sum c_k$, which represents the number of retained parameters in $\boldsymbol{\beta}$, let $\boldsymbol{\beta}^* = \{\beta_k \mid c_k = 1\}$ be a $R \times 1$ vector, which includes all retained parameters in $\boldsymbol{\beta}$, and let $\mathbf{X}^*$ be a $N \times R$ matrix. Obviously, $\mathbf{X}^*$ is a sub-matrix of $\mathbf{X}$, which is obtained by deleting all the columns whose corresponding $c_k = 0$. Simply, we check all the entries of $\mathbf{c}$, if $c_k = 0$, then delete the $k$th column of $\mathbf{X}$, $k = 1, 2, \ldots, N_p$. In the reduced-size model of QTLs mapping, it is obvious that $\mathbf{Y}^*$ is equal to $\mathbf{Y}$. Now the model (3) is simplified now as $\mathbf{Y}^* = \mathbf{X}^*\boldsymbol{\beta}^* + \mathbf{e}$, and can be solved by the standard VLR algorithm again.

In some specific cases, some parameter values are not 0, and can be obtained in advance or learned by other methods. If we want to simplify the model by removing these elements with fixed nonzero values, $\mathbf{Y}^* = \left(y_1^*, y_2^*, \ldots, y_N^*\right)^T$ can be calculated as following

$$y_k^* = y_k - \sum_{j=1}^{N_p} \beta_j x_{kj} \left(1 - c_j\right). \tag{13}$$

### 2.4. The RCVLR algorithm

The iteration of RCVLR includes an inner loop and an outer loop. The inner loop is to solve the linear model with variational linear regression (VLR), and calculate the difference of inferred parameters between two successive iterations to determine whether or not to stop the inner loop. In the outer loop, after getting solution from the inner loop, we delete some variables from the model according to some rules; form the reduced-size model for the next inner loop.

For the outer loop, we can determine whether or not to stop the iteration by comparing the sizes of the two consecutive reduced-size models. If the two sizes are the same, then we stop the outer loop and the whole inference.

It should be noted that for

The detail algorithm is listed as follow:

---

*Input QTLs matrix $\mathbf{X}$, and phenotypic data $\mathbf{Y}$; set initial hyper parameters $a_0 = 1, b_0 = 1$. Set a small value to $\xi$, which is used to judge whether the inner loop comes to an end or not. Set a small value to $\eta$, which is the initial threshold for the first model deletion. Set $\gamma$ as the step length to increase the deletion threshold gradually.*

*Flag=1;*
*h =0;*
*While Flag (outer loop)*
    *Repeat until convergence (inner loop):*

*1. Calculate* $\mathbf{S}_n$ *and* $\boldsymbol{\mu}_n$ *by (11);*

*2. Calculate* $a_n$ *and* $b_n$ *by (9);*

*3. Calculate* $S = \|\boldsymbol{\mu}_n - \boldsymbol{\mu}_{n-1}\|^2$;

*5 If* $S < \xi$, *then end the inner loop, else go to step 1;*

*End repeat*
*h = h +1;*

*Calculate the deletion threshold by* $\eta_h = \eta_1 + h \times \gamma$;

*Calculate the reduced-size model according to Section 2.3;*
*If the size of the reduced-size model is the same as the last time, Flag=0;*
*End While*

*Output* $\boldsymbol{\mu}_n$ *as the inferred* $\boldsymbol{\beta}$.

---

The robustness of VLR had been proved [15]. Due to the nature that RCVLR is composed of several repetitions of VLR, we think that it is also robust, under the condition that we can remove some redundant variables from the model appropriately in each deletion time.

**3. Simulations and Analysis**

We present four simulations to assess the performances of RCVLR. The first one is to compare the performances of VLR and RCVLR. In the second study, we follow the example analysis in [18], simulate the barley genotype data set [17] to assess RCVLR, and compare it with other methods. The third one compares the performance of RCVLR with other two variational inference methods, using simulated data sets with different sample sizes and different marker numbers. The last one uses several variational methods to infer the main and epistatic effects of QTLs with the simulated data set from Xu [2]. The Xu's simulated data set and the barley data set were downloaded from the website http://www.biometrics.tibs.org/.

Suppose that there are $K$ QTLs in a true model, and $K'$ is the QTLs number inferred by a algorithm, which includes $K_{True}$ QTLs that exist in the true model and $K_{False}$ QTLs that do not existed in the true model, and $K' = K_{True} + K_{False}$. For convenience, we use the term, Power of Detection (PD) to signify the ability of finding correct QTLs, defined PD as $PD = K_{True}/K$, and use False Detection Rate (FDR) to indicate the error ratio caused by false inference of the algorithm, defined as $FDR = K_{False}/K'$.

**3.1 Compare RCVLR with VLR**

We first compare RCVLR with VLR by four simulated models with the number of makers 1000, 2000,

3000, 5000 respectively. Averaged $N_m = 20$ QTLs are created for each model. Each entry in $\mathbf{X}$ takes value from set $\{1,2,3\}$ with corresponding probabilities $\{0.25, 0.5, 0.25\}$ respectively. If a marker is a QTL, its corresponding $\beta_j$ is sampled from a uniform distribution over the interval $(-1\ -0.5) \bigcup (0.5\ 1)$; otherwise $\beta_j$ is set to 0. Each $e_j$ in (3) is sampled from a Gaussian distribution $N(0, 0.1)$. $\mathbf{Y}$ is calculated by (3). In simulations, a marker $j$ is judged as a QTL, if $|\beta_j'| \geq 0.05$.

The performances of basic VLR and RCVLR are shown in Fig.1. The hyper parameters of the Gamma distribution in VLR and RCVLR are $a_0 = 1, b_0 = 1$.

The performance of RCVLR is clearly higher than that of VLR does in the terms of PD and FDR, especially when the scale of model is large. The strategy of dynamically reducing the number of variables is useful to remove large number of markers with no effect from the model. In Fig.1, some data points of FDR of VLR are drawn as zeros, which are actually inferred as Nan (not a number) due to the reason that all coefficients inferred by VLR are close to zeros, so $K_{False}, K_{Ture}$ are both zeros, therefore $FDR = K_{False} / (K_{False} + K_{Ture}) = Nan$.

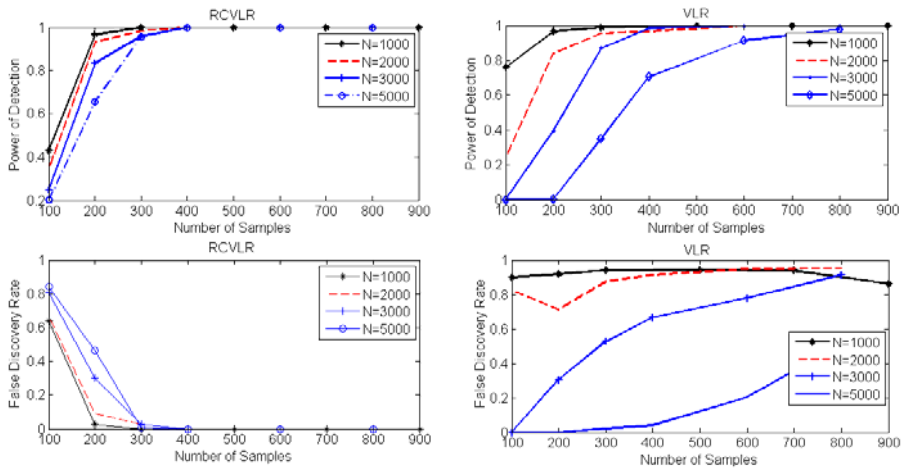

*Figure 1. Performances of RCVLR and VLR for simulated data sets with the size of markers N=1000, 2000, 3000, 5000.*

**3.2 Analysis of barley genotype data**

The barley genotype data set is used as the genotypic data matrix $\mathbf{X}$ in model (3). There are some missing marker genotypes in the original marker dataset, which can be filled by random draws from Bernoulli distribution [18, 20]. We place QTLs exactly on ten markers (1, 10, 20, 30, 40, 50, 60, 70, 80, 90) with effects of (-1, 1, -0.8, 0.8, -0.5, 0.5, -0.3, 0.3, -1.25, 1.25) respectively, and simulate 60 replicated sets of phenotype data $\mathbf{Y}$ with residual variance 0.1. $\mathbf{X}$ and $\mathbf{Y}$ are used to infer QTLs with our RCVLR algorithm, VI-IMP (variational inference with importance sampling) [4], Bayesian adaptive shrinkage (MCMC-BAS) and its variational solution (VB-BAS) [7, 18], Bayesian Lasso (MCMC-BL) and its variational solution (VB-BL) [18, 21], extension Bayesian Lasso (MCMC-EBL) and its variational solution (VB-EBL) [18, 22]. All later six algorithms can be found in [18]. Due to the fact that VB and MCMC approaches do not shrink any marker effect exactly to zero, some criterions should be taken for these algorithms to identify some markers as QTLs. RCVLR can not give us a confidence interval, so we use numerical magnitude judgment: if the effect of one marker is less than 0.05, it is not a QTL, otherwise it is a QTL. In VI-IMP, a marker is judged as a QTL if it has 95% posterior inclusion probabilities included in the model. All other six algorithms use the criterion that a marker is not a QTL, if the credible interval is 95% for the estimated coefficient being zero.

The parameters in algorithms are set as following:

RCVLR: the hyper parameters for the Gamma distribution are $a_0 = 1, b_0 = 1$.

VI-IMP: residual variance $\sigma^2 = [0.08:0.01:0.12]$, the prior variance of the regression coefficients $\sigma_\beta^2 = [0.025:0.025:0.4]$, the (base 10) logarithm of the prior inclusion probability, $\log_{10}^\pi = [-2.5:0.25:-1]$. The prior sample parameters of the Beta distribution for the prior inclusion probability are $a = 0.02, b = 1$.

For VB-BAS, VB-BL, VB-EBL three algorithms, the stopping thresholds and the maximum iteration numbers are all set to 10^-6 and 2000 respectively. For MCMC-BAS, MCMC -BL, MCMC -EBL three algorithms, the total sample numbers, the burn-in numbers and the maximum iteration numbers are all set to 15000, 5000 and 1000 respectively. Additionally, hyper prior $\lambda^2 \sim Ga(1, 0.0001)$ are set in VB-BL and MCMC-BL, $\delta^2 \sim Ga(1, 0.0001), \eta_j^2 \sim Ga(1, 0.0001)$ in VB-EBL and MCMC-EBL [18].

The results are shown in Table 1. We find that: (1) The PD performances of these algorithms are relatively good, their values equal 1 except that VI-IMP's PD is 0.988; (2) there are wide gaps among FDRs of these algorithms. RCVLR has better performance than VB-BL, VB-BAS do, and has similar performance as VB-EBL does; (3) all variational approximation methods' running times are lower than those of MCMC methods. The VI-IMP is much slower than other variational methods. The reason is that, VI-IMP sets a group of combinations of three hyper prior parameters; for each combination in the group, variational algorithm is run to infer the parameters, and then the weighted average is calculated as the final inferred results.

Together, it may suggests that the performance of VB-EBL is the best one among these eight methods, but we will see in the second example that the performance of VB-EBL begin to deteriorate in the case of dealing with large-scale linear models, and RCVLR outperforms the VB-EBL in both PD and FDR. Note that in QTLs mapping study, especially when epistatic effects are considered, the number of variables is usually quite large.

Table 1. Performances of eight variational methods

| Method | MSE | PD | FDR | Runtime(s) |
|---|---|---|---|---|
| RCVLR | 11.145 | 1 | 0.06 | 0.181 |
| VI-IMP | 13.9347 | 0.988 | 0.11 | 31.98 |
| VB-BAS | 9.4541 | 1 | 0.367 | 0.884 |
| MCMC-BAS | 12.3498 | 1 | 0.0004 | 61.25 |
| VB-BL | 7.62 | 1 | 0.368 | 0.13 |
| MCMC-BL | 4.6568 | 1 | 0.212 | 63.84 |
| VB-EBL | 10.9138 | 1 | 0.0138 | 0.167 |
| MCMC-EBL | 8.7219 | 1 | 0.099 | 59.47 |

*MSE represents mean squared error; PD, the power of detection; FDR, the false detection rate.*

### 3.3 Analysis simulated data sets with different sample sizes and marker sizes

In this example, we compare RCVLR with other two variational Bayesian methods, VB-EBL and VI-IMP to study the performance of these algorithms in different sample sizes and marker sizes.

We simulate two types of models, one with 1000 markers, the other one with 5000 markers. The simulation procedure of the data sets for the two models is the same as the one described in section 3.1.

In simulations of RCVLR, a marker $j$ is judged as a QTL if $|\beta_j'| \geq 0.05$; in VI-IMP, the criterion is that it has 95% credible interval included in the model.

Simulation results for the setups described above are shown in Fig.2 (for p=1000) and Fig.3 (for p=5000). It shows that the RCVLR has a better performance than the VB-EBL and the VI-IMP algorithms do in term of PD. The RCVLR and the VI-IMP algorithms have close performance in term of FDR. When the sample size is small, due to the reason that all coefficients inferred are close to zero, VB-EBL can not find any QTL for the model, thus the PD is zero and the FDR is undefined. That is the reason why there is no data point showed in the two FDR graphs for the VB-EBL algorithm. Actually, for the size of markers p=1000, the FDRs of VB-EBL are (Nan Nan) for sample size (100 200) respectively; for p=5000, the FDRs are (Nan Nan) for sample size (100 300).

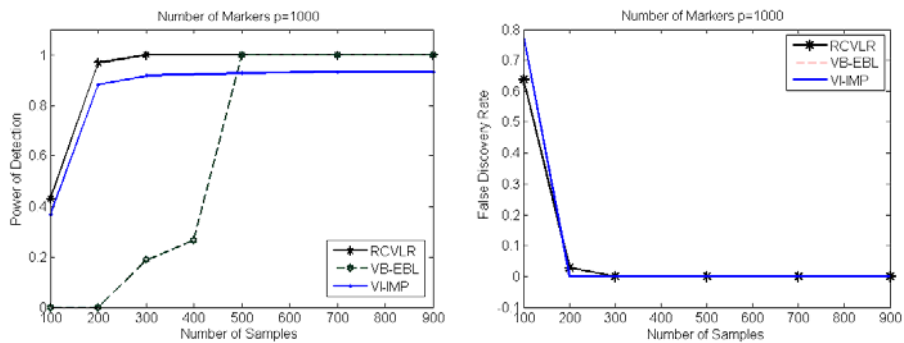

*Figure2. Performances of RCVLR, VB-EBL, VI-IMP for simulated data sets with a size of 1000 markers*

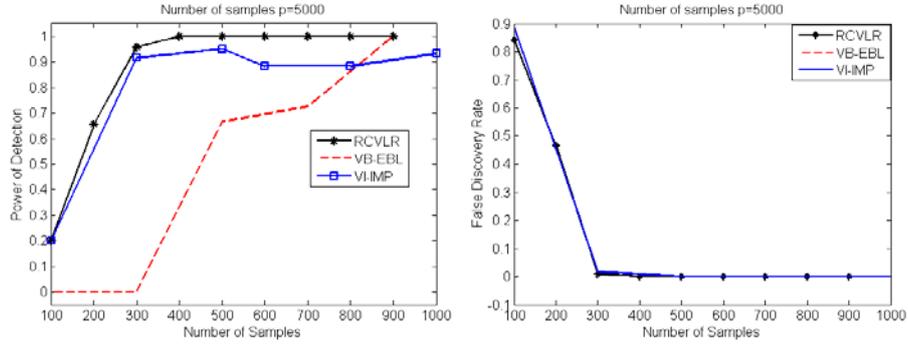

*Figure 3. Performances of RCVLR, VB-EBL, VI-IMP for simulated data sets with a size of 5000 markers*

In the above simulations, the parameters are $a_0=1, b_0=1$. In order to test the impact of hyper parameters for the performances of RCVLR, we run the algorithm under different hyper parameter combinations. The setting is 1000 markers, 20 QTLs. Noise is sampled from $N(0, 0.2)$. The results are shown in Table 2 and Table 3.

Table 2. PD performance of RCVLR with different hyper parameters $(a_0, b_0)$

| (a$_0$,b$_0$)<br>Number of Samples | （0.1,0.1） | （0.1,0.5） | （0.1, 1） | （0.5,0.5） | （0.5,1） | （1, 1） |
|---|---|---|---|---|---|---|
| 250 | 0.9715 | 0.9820 | 0.9894 | 0.9853 | 0.9952 | 0.987 |
| 300 | 0.9762 | 0.9875 | 0.9975 | 1 | 0.998 | 1 |
| 350 | 0.9947 | 1 | 1 | 1 | 1 | 1 |

Table 3. FDR performance of RCVLR with different hyper parameters $(a_0, b_0)$

| (a$_0$,b$_0$)<br>Number of Samples | （0.1,0.1） | （0.1,0.5） | （0.1, 1） | （0.5,0.5） | （0.5,1） | （1, 1） |
|---|---|---|---|---|---|---|
| 250 | 0.0182 | 0.0042 | 0.0280 | 0.0153 | 0.0167 | 0.0150 |
| 300 | 0.0116 | 0.0087 | 0.0108 | 0 | 0.0138 | 0.0143 |
| 350 | 0.0108 | 0 | 0 | 0 | 0.0056 | 0.0131 |

From the two tables above, we see that both the performance of PD and FDR are close under various combinations of hyper parameters. No significant change occurs in PD and FDR under one same simulation setting. For all simulated and real data set in our project, we get good performance with the setting $a_0=1, b_0=1$. Thus we can safely say that RCVLR is not sensitive to the selection of hyper parameters.

### 3.4 Estimate the epistatic effects

Here, we use the backcross data set created by Xu [2] to test the RCVLR performance. The data set has 600 individuals and 121 markers. These makers are evenly distributed through a single chromosome of 1800 cM, and 9 markers are QTLs with main effects, 13 marker pairs have interaction effects. In simulation, the variance of residual errors is 10, and the population mean is 5. We center the marker data and phonetic data and use (3) as the mode in this study. The model contains a total of $121+\binom{121}{2}=7381$ possible effects.

We run the RCVLR and the VI-IMP algorithms to estimate the main and epistatic effects with the simulated data set described above, and compare the results of VB-BAS, VB-BL, VB-EBL provided in [18]. All results are shown in Table 4. The parameter setting of RCVLR and VI-IMP in these algorithms are the same as section 3.1, except that the residual variance $\sigma^2=[8:1:12]$ in VI-IMP. The code for VI-IMP can be downed from //github.com/pcarbo/Variational-inference-for-Bayesian-variable-selection, and the codes for VB-BAS, VB-BL, VB-EBL are available at //www.genetics.org/content/suppl/2011/10/31/genetics.111.134866.DC1.

These algorithms can detect almost all the correct loci with main and epistatic effects. But some loci are failed to be found by some algorithms, such as VB-EBL, VB-BL, RCVLR, VI-IMP failed to find marker pairs (41,61), VB-EBL, VB-BL, VI-IMP failed to find marker pairs (21,22). RCVLR fails to find one main effect at maker (111,111). The effect of pair (20,46) shown in Table 4 is not the true estimated QTL by RCVLR, but is found at the neighboring pair (20, 45), while (56,91) pair's effect is found at (56,90).

Table 4: Estimation main and epistatic effects with simulated data set by several algorithms

| Makers | True effects | RCVLR | VB-IMP | VB-BAS | VB-BL | VB-EBL |
|---|---|---|---|---|---|---|
| (1,1) | 4.47 | 4.4720 | 4.4452 | 4.5419 | 2.7849 | 4.4240 |
| (21,21) | 3.16 | 3.0879 | 3.1616 | 3.1603 | 0.8826 | 3.1502 |
| (31,31) | 2.24 | 2.2473 | 2.2040 | 2.3097 | 0.6063 | 2.2495 |
| (51,51) | 1.58 | 1.2783 | 1.3735 | 1.3123 | 0.0530 | 1.3090 |
| (71,71) | 1.58 | 1.5042 | 1.5259 | 1.7562 | 0.0614 | 1.5042 |
| (91,91) | 1.10 | 0.8604 | 0.8859 | 1.0180 | 0.0233 | 0.8849 |
| (101,101) | 1.10 | 1.3072 | 1.2073 | 1.2624 | 0.0727 | 1.1819 |
| (111,111) | 0.77 | 0 | 0.7429 | 0.4752 | 0.0169 | 0.6737 |
| (121,121) | 0.77 | 0.6963 | 0.6915 | 0.4799 | 0.0490 | 0.5565 |
| (1,11) | 1.00 | 0.7797 | 0.7503 | 0.6421 | 0.0199 | 0.6545 |
| (2,119) | 3.87 | 3.8586 | 3.8723 | 3.4860 | 1.0283 | 3.8556 |
| (10,91) | 1.30 | 1.2473 | 1.3438 | 1.1521 | 0.0432 | 1.2087 |
| (15,75) | 1.73 | 1.4988 | 1.5571 | 1.2712 | 0.0810 | 1.5027 |
| (20,46) | 1.00 | 0.9011* | 0.7996 | 0.7548* | 0.0200 | 0.8834 |
| (21,22) | 1.00 | 0.3416 | 0 | 0.7588 | 0.0139 | 0 |
| (26,91) | 1.00 | 1.1965 | 1.1643 | 1.2658 | 0.0449 | 1.2263 |
| (41,61) | 0.71 | 0 | 0 | 0.3119* | 0.0038 | 0 |
| (56,91) | 3.16 | 2.9931* | 3.0597* | 2.8543* | 0.6414* | 3.0348* |
| (65,85) | 2.24 | 2.4868 | 2.4490 | 2.2781 | 0.3207 | 2.4204 |
| (86,96) | 0.89 | 0.9772 | 1.0489 | 0.8318 | 0.0211 | 0.9601 |
| (101,105) | 1.00 | 0.8778 | 1.0492 | 1.0368 | 0.0516 | 0.9754 |
| (111,121) | 2.24 | 2.3706 | 2.3345 | 2.0512 | 0.4615 | 2.3373 |

*Means that the estimated QTL effects are not in the position of the true QTL, but in their neighboring markers.

### 4. Conclusions and discussion

The main goal of this paper is to propose an improved variational linear regression approach for high scale variable selection problems such as the ones arising in epistatic analysis. Variational approximation is a classic method to solve linear model, but its accuracy is limited in the case of there are many more variables than samples. Several improved methods were proposed for the variational approximation, such as hierarchical shrinkage, importance sampling and so on. But these methods still lack of efficiency or accuracy. We proposed a simple but effective method to improve the performance of VLR by dynamically deleting some specific variables, which can greatly improve the accuracy of VLR.

The reason that RCVLR is more accurate than VB-EBL, VB-BAS, VB-BL may lie in that it can derive the approximate marginal distributions of all variables in a single process, while hierarchical shrinkage infers each variable separately, therefore some correlation information among variables is lost. The VI-IMP considers the correlation among variables, but it uses important sampling and a weighted average method to improve the accuracy; therefore the computational cost is higher than other variational approaches.

RCVLR can obtain good performance in general, and is not sensitive to the choice of hyper parameters, while some other variational approaches, such as VB-EBL, rely heavily on selecting appropriate parameters to get good performance [18]. VI-IMP averages the inferred results under different prior combinations. But how to determine the prior ranges is a problem unsolved [4]. If an inappropriate range is inputted, the performance may be poor. But if a very wide range is set to cover the exact parameters, the algorithm will need many more samples and thus need much more computational time.

**Acknowledgments**

This work was jointly supported by The National Natural Science Foundation of China (No.61271207, No.61174013), and the Jiangsu Overseas Research \& Training Program for University Prominent Young \& Middle-aged Teachers and Presidents.